\documentclass[aps, prl, reprint, nofootinbib]{revtex4-2}
\usepackage{amsmath}
\usepackage{amsfonts}
\usepackage{amssymb}
\usepackage{graphicx}
\usepackage{mathrsfs}
\usepackage{slashed}
\usepackage{xcolor}	
\usepackage{subcaption}
\usepackage{relsize}

\newcommand{\e}{\textrm{e}}

\newcommand{\symb}{\textrm{symb}}

\newcommand{\jhat}[1]{\hspace{0.3em}\widehat{\hspace{-0.4em}#1\hspace{-0.4em}}\hspace{0.4em}}

\def\bear{\begin{eqnarray}}
\def\ear{\end{eqnarray}}
\def\e{{\rm e}}

\newcommand{\ds}{\slashed{\partial}}
\newcommand{\dsp}{\ds^{\prime}}
\newcommand{\Ds}{\slashed{D}}
\newcommand{\Dsp}{\Ds^{\prime}}
\newcommand{\Dsph}{\jhat{\Ds}{}^{\prime}\!}

\newcommand{\As}{\slashed{A}}
\newcommand{\Js}{\slashed{J}}

\newcommand{\be}{\begin{equation}}
\newcommand{\ee}{\end{equation}}
\newcommand{\bi}{\begin{itemize}}
\newcommand{\ei}{\end{itemize}}
\newcommand{\bea}{\begin{eqnarray}}
\newcommand{\eea}{\end{eqnarray}}

\begin{document}

\title{Non-perturbative gauge transformations of arbitrary fermion correlation functions in quantum electrodynamics}


\author{José Nicasio}
\email{jose.nicasio@umich.mx}
\affiliation{Instituto de F\'isica y Matem\'aticas
Universidad Michoacana de San Nicol\'as de Hidalgo
Edificio C-3, Apdo. Postal 2-82
C.P. 58040, Morelia, Michoac\'an, M\'exico}
\author{Naser  Ahmadiniaz}
\email{n.ahmadiniaz@hzdr.de}
\affiliation{Helmholtz-Zentrum Dresden-Rossendorf, Bautzner Landstra\ss e 400, 01328 Dresden, Germany}
\author{James P. Edwards}
\email{jedwards@ifm.umich.mx}
\affiliation{Instituto de F\'isica y Matem\'aticas
Universidad Michoacana de San Nicol\'as de Hidalgo
Edificio C-3, Apdo. Postal 2-82
C.P. 58040, Morelia, Michoac\'an, M\'exico}
\author{Christian Schubert}
\email{schubert@ifm.umich.mx}
\affiliation{Instituto de F\'isica y Matem\'aticas
Universidad Michoacana de San Nicol\'as de Hidalgo
Edificio C-3, Apdo. Postal 2-82
C.P. 58040, Morelia, Michoac\'an, M\'exico}


\begin{abstract}
We study the transformation of the dressed electron propagator and the general $N$-point functions under a change in the covariant gauge of internal photon propagators. We re-establish the well known Landau-Khalatnikov-Fradkin transformation for the propagator and generalise it to arbitrary correlation functions in configuration space, finding that it coincides with the analogous result for scalar fields. We comment on the consequences for perturbative application in momentum-space.
\end{abstract}

\maketitle

\section{Introduction}
\label{secIntro}
Obtaining non-perturbative information about the structure of correlation functions in quantum electrodynamics (QED) continues to be a difficult aspect of quantum field theory, yet remains a subject of intense investigation. The gauge dependence of various quantities has received particular attention, such as for the decomposition of the QED vertex \cite{BallChiu} whose coefficients have been determined to one-loop order in various gauges \cite{YFVertex, KizVertex} and in diverse analyses of the one- \cite{Davydychev, GluonVertex1} and two-loop \cite{VertexQCD1, VertexQCD2} quark-gluon vertex  (see \cite{QED31, QED32, QED33,QED34} for similar results in three-dimensions, QED$_{3}$, \cite{ScalarVertex1, ScalarVertex2} in scalar QED or \cite{RQED1, RQED2} for reduced QED). The three- and four-gluon vertices have also been calculated, off-shell, in covariant gauges up to two-loop order \cite{BallChiuQCD1,GluonVertex1, GluonVertex2, GluonVertex3, GluonVertex4,Davydychev:1998aw, GluonVertex5, GluonVertex6, GluonVertex7, GluonVertex8, GluonVertex9, GluonVertex10, GluonVertex11,Papavassiliou:1992ia,Binosi:2014kka,cyrol2015dyson}, relevant for studies using the Dyson-Schwinger equations and of infrared divergences within quantum chromodynamics (QCD) --  see \cite{HUBER20201} for a review of such non-perturbative analyses.

Despite the gauge invariance of physical observables such as cross sections, the $N$-point Green functions of a gauge theory depend strongly on the gauge of internal gauge bosons (below we recall that the transformations of external photons or gluons are fixed by Ward-Takahashi \cite{WTI} or Slavnov-Taylor \cite{STI1, STI2} identities). Since various gauge choices are used in perturbative studies, amongst them Feynman gauge, Landau gauge or Yennie-Fried gauge \cite{YFG}, there is great theoretical interest in the variation of such quantities under a change in gauge. The transformation of the QED $2$-point function -- the propagator -- and the vertex operator under changes in the covariant gauge parameter were initially examined by Landau and Khalatnikov \cite{LandauLKFT} and independently by Fradkin \cite{FradkinLKFT}, later analysed using functional methods in \cite{ZuminoLKFT1, ZuminoLKFT2}. 

We summarise these fully non-perturbative ``LKF'' transformations that apply in configuration space. The fermion propagator in covariant gauge with parameter $\xi$ is denoted by $S(x; \xi)$ and effecting a variation in the gauge $\xi \rightarrow \xi + \Delta \xi$ induces the transformation
\begin{equation}
	S(x; \xi + \Delta \xi) = S(x; \xi) \e^{-i\Delta \xi\left[\Delta_{D}(x ) - \Delta_{D}(0)\right]}\, ,
	\label{eqLKFTProp}
\end{equation} 
with $\Delta_{D}(x)$ a function corresponding to the gauge fixing \cite{LandauLKFT} (see (\ref{eqDeltaD})). Since the exponent is order $\alpha$, the fine structure constant, it is clear that from the propagator evaluated to a given loop order the transformation can construct gauge dependent parts of higher-loop diagrams \cite{AdnanProc, AdnanRaya}. Furthermore, the LKF transformations relate the wave function normalisation constants in different gauges \cite{ZuminoLKFT1} (perturbative application of the transformations can be found in \cite{Kissler, KissDirk}).  Despite containing potentially significant information about the structure of gauge theory interactions, these transformations remain less well analysed than the constraints implied by Ward-Takahashi / Slavnov-Taylor identities \cite{Aguilar}.

There is substantial interest, in particular in QCD and extensions of the Gribov-Zwanziger scenario \cite{gribov1978quantization,zwanziger1989local}, in LKF transformations for $N$-point functions \cite{Pietro1}. Recent methods have successfully derived the transformation rules for the gluon propagator by including auxiliary fields and exploiting a BRST invariance \cite{Sonoda, Pietro2, Ply1, Ply2}.

The LKF transformations for $N$-point correlation functions in scalar QED were recently uncovered in \cite{LKFTWL1, LKFTWL2} using the alternative \textit{worldline formalism} of quantum field theory \cite{Strassler1, ChrisRev, UsRev, WLNotes}. They are fully determined by the ordered (quenched) ${(N = 2n)}$-point function where $n$ fields $\phi(x_{i})$ are paired with $n$ conjugate fields $\phi^{\dagger}(x^{\prime}_{\pi(i)})$ for ${\pi \in S_{n}}$, denoted in covariant gauge with parameter $\xi$ by $\mathcal{A}_{\pi}(x_{1}, \ldots , x_{n}; x^{\prime}_{\pi(1)},\ldots, x^{\prime}_{\pi(n)} | \xi)$, which transforms as
\begin{align}
	&\mathcal{A}_{\pi}(x_{1}, \ldots , x_{n}; x^{\prime}_{\pi(1)},\ldots, x^{\prime}_{\pi(n)} | \xi + \Delta \xi)	\label{eqScalarLKFT}\\
	&= \prod_{k, l = 1}^{n}\e^{-\Delta_{\xi}S_{i \pi}^{(k, l)}} \mathcal{A}_{\pi}(x_{1}, \ldots , x_{n}; x^{\prime}_{\pi(1)},\ldots, x^{\prime}_{\pi(n)} | \xi )\,. \nonumber
\end{align}  
This is the direct generalisation of (\ref{eqLKFTProp}) given in terms of functions $\Delta_{\xi}S_{i \pi}^{(k, l)}$ later defined in (\ref{eqDeltaXis}). This Letter reports on the extension of this result to the  physical case of spinor QED to obtain the complete LKF transformation of the $N$-point fermionic Green functions that enter the calculation of scattering amplitudes in perturbation theory. A longer communication will supply more calculational details and describe various applications \cite{UsLKFT}.

\subsection{Overview}
\label{secOver}
We use the worldline representation of the spinor propagator that has only recently been given in \cite{fppaper1, fppaper2,Corradini:2020prz}, based upon the second order formalism of Dirac fermions \cite{SO1, SO2}, which in a background field with gauge potential $\textrm{A}(x) = A_{\mu}(x)dx^{\mu}$ is decomposed as
\begin{equation}
	S^{x'x}[A] = \big[m + i\Dsp\big]K^{x'x}[A]\,, 
	\label{eqSProp}
\end{equation}
where we define the covariant derivative by ${D^{\prime}_{\mu}(x) = \partial^{\prime}_{\mu} + ieA_{\mu}(x')}$ and $K^{x'x}[A]$ is a matrix-valued kernel defined below. Using the background field method \cite{Abbott1, Abbott2, Abbott3} we split the gauge field $\textrm{A} = \textrm{A}^{\gamma} + \bar{\textrm{A}}$ into a part $\textrm{A}^{\gamma}$ associated to external photons and a ``quantum'' piece $\bar{\textrm{A}}$ which generates virtual photon propagators evaluated in a chosen covariant gauge.

Writing the ``backgroundless'' part of the propagator as $S_{0}^{x'x}[A ; \xi] =  \big[m + i\big(\ds^{\prime} + ie\As^{\gamma}\big)\big]K^{x'x}[A^{\gamma} + \bar{A}],$ we show that the \textit{backgroundless} (ordered) $(N=2n)$-point functions defined below transform as (again $\pi \in S_{n}$)
\begin{align}
	\hspace{-0.5em}&\mathcal{S}_{0\pi}(x_{1}, \ldots , x_{n}; x^{\prime}_{\pi(1)},\ldots, x^{\prime}_{\pi(n)} |  \xi + \Delta \xi) =\label{eqS0LKFT} \\
	 \hspace{-0.5em} &\Big< \prod_{i = 1}^{n}\big[m + i\big(\ds^{\prime}_{i} + ie \As^{\gamma}\big)\big]K^{x^{\prime}_{\pi(i)} x_{i}}[A^{\gamma}\! + \!\bar{A} ]\Big>_{\bar{A}, \,\xi} \!\prod_{k, l = 1}^{n}\e^{-\Delta_{\xi}S_{i \pi}^{(k, l)}} \nonumber 
\end{align}
where derivatives $\ds^{\prime}_{i} := \gamma^{\mu}\partial_{x_{i}^{\prime \mu}}$ act through onto the \textit{same} scalar factor as in (\ref{eqScalarLKFT}).  The additional factors of $\bar{\As}$ in the complete correlation functions cancel these unwanted derivatives to provide the generalised LKF transformations of the full (ordered) fermionic $N$-point functions,
\begin{align}
	&\mathcal{S}_{\pi}(x_{1}, \ldots , x_{n}; x^{\prime}_{\pi(1)},\ldots, x^{\prime}_{\pi(n)} | \xi + \Delta \xi ) 	\label{eqSLKFT} \\ 	
	  =&\prod_{k, l = 1}^{n}\e^{-\Delta_{\xi}S_{i \pi}^{(k, l)}} \mathcal{S}_{\pi}(x_{1}, \ldots , x_{n}; x^{\prime}_{\pi(1)},\ldots, x^{\prime}_{\pi(n)} | \xi )\, , \nonumber
\end{align}
in which the exponential factor is identical to the scalar result of \cite{LKFTWL1, LKFTWL2}. In the following sections we prove these claims, giving further details in the appendix.

\section{Gauge transformations of Green functions}
\label{secLKFT}
The photon propagator corresponding to covariant gauge parameter $\xi$, $G_{\mu\nu}(x - x^{\prime}; \xi) := \langle \bar{A}_{\mu}(x)\bar{A}_{\nu}(x^{\prime})\rangle_{\xi}$, takes the form
\begin{equation}
	G_{\mu\nu}(x - x^{\prime}; \xi) = G_{\mu\nu}(x - x^{\prime} ; \hat{\xi}) + \frac{\Delta \xi}{e^{2}}  \partial_{\mu}\partial_{\nu}\Delta_{D}(x - x^{\prime})
	\label{eqGPhoton}
\end{equation}
where $\hat{\xi}$ refers to an arbitrary reference covariant gauge and $\Delta \xi = \xi - \hat{\xi}$. Here $\Delta_{D}(y)$ fixes the non-physical longitudinal part of the propagator \cite{AdnanRaya, LKFTWL1}, 
\begin{equation}
	\Delta_{D}(y) =  -\frac{ie^{2}(\mu)}{16\pi^{\frac{D}{2}}}\Gamma\Big[\frac{D}{2} - 2\Big](\mu y)^{4-D}
	\label{eqDeltaD}
\end{equation}
where  $e^{2} \rightarrow \mu^{4-D}e^{2}(\mu)$ defines the usual arbitrary mass scale. The momentum space transformation of ${\bar{A}_{\mu}(k) \rightarrow \bar{A}_{\mu}(k) -ik_{\mu}\phi(k)}$ change the longitudinal part of the photon two-point function, (\ref{eqGPhoton}), according to
\begin{equation*}
	\langle \bar{A}_{\mu}(k)\bar{A}_{\nu}(-k)\rangle_{\xi} \rightarrow  \langle \bar{A}_{\mu}(k)\bar{A}_{\nu}(-k)\rangle_{\xi} - \xi \frac{k_{\mu}k_{\nu}}{k^{4}}.
\end{equation*}
Through (\ref{eqLKFTProp}), the function $\Delta_{D}$ transforms the propagator between covariant gauges, as developed in $3$- and $4$-dimensional space-time for the cases of spinor \cite{AdnanRaya} and scalar \cite{ScalarPerturb} QED. As is clear in the original derivations and confirmed here, the LKF transformation of the propagator coincides for scalar and spinor QED, now extended to arbitrary correlation functions in this Letter. 

\subsection{Correlation functions in spinor QED}
\label{secCorrelation}
We follow the techniques in \cite{LKFTWL1, LKFTWL2} using the first quantised representation of the fermionic correlation functions developed in \cite{fppaper1, fppaper2}. The Dirac propagator in a background electromagnetic field $\textrm{A} = A_{\mu}dx^{\mu}$ in the second order formalism is the matrix element
\begin{equation*}
	S_{\beta, \alpha}^{x'x}[A] := \langle x', \beta | [m - i \Ds]^{-1} | x, \alpha \rangle = \big[m+i\Dsp\big]_{\beta \sigma}K^{x'x}_{\sigma \alpha}[A]\, .
\end{equation*}
The \textit{kernel}, $K^{x'x}_{\sigma \alpha}$, admits the path integral representation 
\begin{align}
	\hspace{-0.75em}K^{x'x}&[A]=2^{-\frac{D}{2}} \symb^{-1} \hspace{-0.5em}\int_{0}^{\infty}  \hspace{-0.5em}dT e^{-m^{2}T}  \!\int_{x(0) = x}^{x(T) = x'}\hspace{-1.5em}\mathscr{D}x\int_{\psi(0) + \psi(T) = 0}\hspace{-3em}\mathscr{D}\psi \, \nonumber \\
	\hspace{-0.75em}&\e^{-\int_{0}^{T}d\tau\, \left[ \frac{\dot{x}^{2}}{4} + \frac{1}{2} \psi \cdot \dot{\psi} +ie \dot{x} \cdot A(x) -ie (\psi + \eta) \cdot F(x) \cdot (\psi + \eta)\right]},
	\label{eqKPathInt}
\end{align}
over trajectories from $x$ to $x'$ and anti-periodic Grassmann variables $\psi^{\mu}(\tau)$ (that generate the ``Feynman spin factor'' \cite{FeynSp}). The ``symbol map'' acts on the constant Grassmann variables $\eta^{\mu}$ as
\begin{equation}
	\symb\left\{ \gamma^{[\mu_{1}}\cdots \gamma^{\mu_{n}]} \right\} \equiv (-i\sqrt{2})^{n}\eta^{\mu_{1}}\ldots \eta^{\mu_{n}},
\end{equation}
where we anti-symmetrise with the appropriate combinatorial factor. We shall see that the evaluation of this path integral is not important here, except for contributions from the action at the endpoints of the worldlines.

In position space the ${(N=2n)}$-point function is decomposed into $n!$ partial amplitudes $\mathcal{S}(x_{1},\ldots, x_{n}; x^{\prime}_{1}, \ldots x^{\prime}_{n} | \xi )= \sum_{\pi \in S_{n}} \hspace{-0.25em}\mathcal{S}_{\pi}(x_{1},\ldots, x_{n}; x^{\prime}_{\pi(1)}, \ldots x^{\prime}_{\pi(n)} | \xi)$ where in $\mathcal{S}_{\pi}$ the spinor field $\Psi(x^{\prime}_{\pi(i)})$ is connected to the conjugate field $\bar{\Psi}(x_{i})$. The gauge potential is then split into plane waves, $A^{\gamma}_{\mu}(x) = \sum_{i = 1}^{N}\varepsilon_{i\mu}e^{ik_{i}\cdot x},$ and a quantum background field, $\bar{A}$, such that internal photons are produced by functional integration over $\bar{A}$:
\begin{align}
	 \mathcal{S}_{\pi}&(x_{1},\ldots, x_{n}; x^{\prime}_{\pi(1)}, \ldots x^{\prime}_{\pi(n)} | \xi) 	 \label{eqSpartial} \\
	 &=\Big \langle \prod_{i = 1}^{n}\big[m + i \Dsp_{i}]K_{i}^{x'_{\pi(i)}x_{i}}[A^{\gamma}+ \bar{A}]\Big\rangle_{\bar{A}, \,\xi}\,, \nonumber
\end{align}
with the expectation values of $\bar{A}$ in accordance with (\ref{eqGPhoton}).

\subsection{Dependence on the gauge parameter}
\label{secGauge}
As in \cite{LKFTWL1, LKFTWL2, fppaper1}, the Ward identity shows that gauge transformations of external photons do not affect on-shell matrix elements\footnote{The plane wave decomposition leads to insertions of vertex operators, \begin{equation*}\hspace{0.5em}V_{\eta}^{x'x}[k,\varepsilon] := \int_{0}^{T}d\tau \big[ \varepsilon \cdot \dot{x} - i(\psi + \eta) \cdot f \cdot (\psi + \eta)\big]\e^{i k \cdot x},\end{equation*} where $f_{\mu\nu}:= 2k_{[\mu}\varepsilon_{\nu]}$ is the (invariant) photon field strength tensor, under the path integral. Under $\varepsilon \rightarrow \varepsilon + \xi k$ the vertex operator picks up a term $-i\xi \big(\e^{i k \cdot x'} - \e^{i k \cdot x}\big)$ which no longer has the required LSZ pole structure to contribute on-shell.}, so we henceforth ignore their gauge variation. For virtual photons, we define a path integral over $\bar{A}$  with the gauge fixed Maxwell action $S(\xi) = \int d^{4}x [-\frac{1}{4}\bar{F}_{\mu\nu}\bar{F}^{\mu\nu} - (\partial \cdot \bar{A})^{2}/(2\xi)]$ that is Gaussian. The insertions of $\bar{A}$ in the propagator prefactors can be generated by functional differentiation of a source term $\e^{ie \int d^{D}x J(x)\cdot \bar{A}(x)}$  appended to the worldline action.

These considerations motivate us to consider the gauge transformation of the expectation value (for $\pi \in S_{M}$)
\begin{equation}
	\mathcal{I}[J, M; \xi):= \Big \langle \e^{ie \int d^{D}x J(x)\cdot \bar{A}(x)}\prod_{j=1}^{M}K_{j}^{x'_{\pi(j)}x_{j}}[A^{\gamma}+ \bar{A}]\Big\rangle_{\bar{A}, \,\xi}\,.
\end{equation}
Using the path integral representation (\ref{eqKPathInt}) and carrying out the integral over $\bar{A}$  leads to a function inserted under said path integrals (further details in the appendix) which depends upon the gauge parameter, $\xi$ according to
\begin{equation}
	\hspace{-0.225em} \mathcal{I}[J, M; \xi + \Delta \xi) = \mathcal{I}[J,M; \xi)\,\e^{-\sum_{k, l=1}^{M}\Delta_{\xi}S_{i\pi}^{(k, l)} + \Delta_{\xi}I_{M}},
	 \label{eqIM}
\end{equation}
where we defined the variations
\begin{widetext}
\begin{align}
	\hspace{-1em}\Delta_{\xi}I_{M} &= \frac{\Delta \xi e^{2}}{16 \pi^{\frac{D}{2}}}\Gamma\Big[\frac{D}{2} - 2\Big] \hspace{-0.25em}\left(\sum_{i = 1}^{M} \int_{0}^{T_{i}}\hspace{-0.5em}d\tau_{i}\hspace{-0.25em}\int \hspace{-0.25em}d^{D}x J(x) \cdot \partial_{x} \partial_{\tau_{i}}\big[(x - x_{i})^2\big]^{2-\frac{D}{2}} \hspace{-0.25em}-\frac{1}{2}\iint\hspace{-0.25em} d^{D}x d^{D}x^{\prime} J(x) \cdot \partial_{x} J(x^{\prime})\cdot \partial_{x^{\prime}}\big[(x - x^{\prime})^{2}\big]^{2 - \frac{D}{2}}\right)\nonumber \\
	\hspace{-1em}\Delta_{\xi}S_{i\pi}^{(k, l)} &= \frac{\Delta\xi e^{2}}{32 \pi^{\frac{D}{2}}} \Gamma\Big[\frac{D}{2} - 2\Big]\int_{0}^{T_{k}}\hspace{-0.25em}d\tau_{k} \int_{0}^{T_{l}}\hspace{-0.25em}d\tau_{l} \, \partial_{\tau_{k}}\partial_{\tau_{l}}\left[(x_{k} - x_{l})^{2}\right]^{2- \frac{D}{2}}\, .
	\label{eqDeltaXis}
\end{align}
\end{widetext}
Note that $\Delta_{\xi}S^{(k, l)}_{\pi}$ is precisely the (scalar) variation in the interaction part of the worldline action that arose in \cite{LKFTWL1, LKFTWL2}; the spin degrees of freedom do not modify the form of the LKF transformation of $\mathcal{I}$. Since, being total derivatives, the variations in (\ref{eqDeltaXis}) depend only on the worldline endpoints they can be taken outside of the path integrals which yields (\ref{eqIM}) and implies that the gauge transformation of the correlation functions is fully determined by the variation of the \textit{quenched} amplitudes.

To proceed we must include the covariant derivatives in the prefactor of (\ref{eqSpartial}). Note that the \textit{backgroundless} part of $\mathcal{S}$ (removing $\bar{A}$ from the covariant derivatives), written as $\mathcal{S}_{0}$, transforms similarly to (\ref{eqIM}) without source, except that now the derivatives act through onto the exponential factor on the right hand side: since 
\begin{align}
	&\Big \langle \e^{ie \int d^{D}x J(x)\cdot \bar{A}(x)}\prod_{i = 1}^{K}\dsp_{i}\prod_{j=1}^{M}K_{j}^{x'_{\pi(j)}x_{j}}[A^{\gamma}+ \bar{A}]\Big\rangle_{\bar{A}, \,\xi + \Delta \xi} \nonumber \\
	&\hspace{1.5em}= \Big[\prod_{i = 1}^{K}\dsp_{i}\Big]\Big[ \mathcal{I}[ J, M; \xi)\e^{-\sum_{k, l=1}^{M}\Delta_{\xi}S_{\pi}^{(k, l)} + \Delta_{\xi}I_{M}}\Big]\,, 
	\label{eqDerivI}
\end{align}
putting $J = 0$ leads to the first key result, equation (\ref{eqS0LKFT}).

What is missing, then, is the incorporation of the prefactors $\bar{A}$ in the full covariant derivative. For the propagator ($N=2$), it is straightforward to evaluate 
\begin{equation}
	-e\Big\langle \bar{A}(x')K^{x'x}\big[A^{\gamma} + \bar{A}\big]\Big\rangle_{\bar{A},\, \xi} = i\frac{\delta}{\delta\Js(x')}\mathcal{I}[J, 1; \xi]\Big|_{J = 0},\nonumber
\end{equation}
using (\ref{eqIM}) to determine that a variation in $\xi$ transforms this term to
\begin{equation}
	\hspace{-0.75em}\Big[-e\Big\langle \bar{A}(x')K^{x'x}\big[A^{\gamma} + \bar{A}\big]\Big\rangle_{\bar{A}, \xi} -\Big\langle K^{x'x}\big[A^{\gamma} + \bar{A}\big]\Big\rangle_{\bar{A},\, \xi}\hspace{-0.25em}i\dsp\Big]\e^{-\Delta_{\xi}S_{i}}, 
	\label{eqDeltaAK}
\end{equation}
where we have used that 
\begin{equation}
	\frac{\delta}{\delta \Js(x_{i}^{\prime})} \Delta_{\xi}I_{M}\Big|_{J=0} = \dsp_{i}\Big[ \sum_{k, l = 1}^{M}\Delta_{\xi}S^{(k, l)}_{i\pi} \Big]\, ,
	\label{eqdeltaJ}
\end{equation}
with $M = 1$. Now $\mathcal{S}_{0}(x_{1}; x^{\prime}_{1} | \xi + \Delta \xi)$ defined by setting $N = 2$ in (\ref{eqS0LKFT}) contains the same derivative of $\Delta_{\xi}S$ that is cancelled by the second term in (\ref{eqDeltaAK}) so that in the complete propagator the exponential factor can be taken through, giving 
\begin{align}
	\Big\langle &\big[m + i \Dsp\big]K^{x^{\prime}x}[A^{\gamma} + \bar{A}]\Big\rangle_{\bar{A}, \, \xi + \Delta \xi} \nonumber\\
	&= \e^{-\Delta_{\xi}S_{i}}\Big\langle \big[m + i \Dsp\big]K^{x^{\prime}x}[A^{\gamma} + \bar{A}]\Big \rangle_{\bar{A}, \, \xi }\, .
\end{align}
This is the worldline derivation of the original LKF transformations for the propagator in spinor QED, reproducing (\ref{eqSLKFT}) with $N = 2$ (one readily verifies the exponent matches that of (\ref{eqLKFTProp}) in the conventional notation). 

\subsubsection{$N$-point functions}
\label{secNpoint}
Finally this result must be generalised to an arbitrary number of propagators which we achieve as a special case of analysing the more general functional (for $M \geqslant K$)
\begin{align}
	&\mathcal{J}[J, K, M; \xi):= \label{eqJK} \\
	&\Big\langle \e^{ie \int d^{D}x J(x)\cdot \bar{A}(x)}\prod_{i=1}^{K}\big[m + i \Dsp_{i}\big]\prod_{j=1}^{M}K_{j}^{x'_{\pi(j)}x_{j}}[A^{\gamma}+ \bar{A}]\Big\rangle_{\bar{A}, \,\xi}\,. \nonumber
\end{align}
We prove by induction on $K$ that $\mathcal{J}$ transforms similarly to $\mathcal{I}$ in equation (\ref{eqIM}), up to additional terms involving derivatives of $\Delta_{\xi}I_{N}$ that vanish when $J = 0$. Indeed $\mathcal{I}$ and the calculation that results in (\ref{eqDeltaAK}) correspond to the $K = 0$  and $K = 1$ cases. 

Assuming the result holds for $K = \kappa-1$ we now express $\mathcal{J}[J, \kappa,  M; \xi)$ as 
\begin{align}
	&\Big[m + i\dsp_{1} - e\As^{\gamma}(x_{1}^{\prime})	+i\frac{\delta}{\delta \Js(x_{1}^{\prime})}\Big]\mathcal{J}[J, \kappa-1, M; \xi) \,,
	\label{eqJkappa}
\end{align}
where in $\mathcal{J}$ the product over $i$ runs from $i = 2$ to $i = \kappa $. Varying the gauge parameter we are led to a cancellation similar to that of the $2$-point function: the partial derivative in (\ref{eqJkappa}) acting on the exponent of (\ref{eqIM}) removes the contribution from the functional derivative on the same due to (\ref{eqdeltaJ}). It also generates additional terms that do not survive when $J = 0$ -- the delicate point, detailed in the appendix, is that derivatives of such contributions from the $K = \kappa - 1$ case cannot produce anything new that would survive this limit.

This leaves the result that was to be proved:
\begin{align}
	&\mathcal{J}[J, \kappa, M; \xi + \Delta \xi) \label{eqJxi} \\
	= &\mathcal{J}[J, \kappa, M; \xi)\e^{-\sum_{k, l=1}^{M}\Delta_{\xi}S_{i\pi}^{(k, l)} + \Delta_{\xi}I_{M}}+ \mathcal{F}[J, \kappa, M; \xi) \,, \nonumber 
\end{align}
where the functional $\mathcal{F}$ satisfies $\mathcal{F}[0, K, M; \xi) = 0$. It turns out, then, that the factors of $\bar{A}$ in the propagator prefactors generate terms which cancel derivatives of the LKF exponent in (\ref{eqIM}) to allow it to be factorised (this disproves the conjecture given in the conclusion of \cite{LKFTWL1} that they would modify the transformation for $N > 2$).

The $(N = 2n)$-point ordered correlation function is in fact a specific case of $\mathcal{J}$ with $K = n = M$ and no source (so in particular, $\Delta_{\xi}I_{n} = 0$),
\begin{equation}
	\mathcal{S}(x_{1}, \ldots x_{n}; x^{\prime}_{1}, \ldots x^{\prime}_{n} | \xi) = \mathcal{J}[0, n, n; \xi)\, ,
\end{equation}
and (\ref{eqJxi}) immediately gives our main result, equation (\ref{eqSLKFT}). We can strengthen this by remarking that direct calculation of the sum over $k$ and $l$ in the exponent of (\ref{eqJxi}) reveals that the exponential factor does not depend upon the permutation $\pi$, since the sum includes all possible pairing of initial points $x_{i}$ to final points $x_{i}^{\prime}$  (see also the discussion in \cite{UsLKFT}). Thus, the exponential factorises out of the sum over partial amplitudes leading to the multiplicative transformation for the complete correlator
\begin{align}
	&S(x_1 \ldots x_n ; x'_{1} \ldots x'_{n} | \xi + \Delta \xi) \\
	&= \e^{-\sum_{k, l=1}^{M}\Delta_{\xi}S_{i}^{(k, l)} } S(x_1 \ldots x_n ; x'_{1} \ldots x'_{n} | \xi )\, , \nonumber
\end{align}
which closely resembles the original LKF transformation of the propagator (the exponent can be calculated with any permutation).

\section{Conclusion}
\label{secConc}
Using the first quantised (worldline) representation of spinor QED we have rederived the LKF transformations in position space for the gauge variation of the Dirac propagator and have successfully extended these transformations to the full $N$-point functions. The gauge variation was determined by examining the transformation of the ordered, quenched amplitudes which we have explicitly obtained, adding to earlier working by noting that their LKF transformation coincides for all partial amplitudes. Our result represents an all orders or non-perturbative calculation for arbitrary correlation functions that, of course, reduces to the original LKF transformation for $N = 2$. 

The functional form of the transformation is equivalent to the gauge variation seen for the ordered amplitudes in scalar QED, since a delicate cancellation occurs between derivative terms that could have spoiled the exponential factorisation that is characteristic of the LKF transformations. As a consequence, the perturbative application of the transformations will carry through analogously to the results in \cite{LKFTWL1, LKFTWL2}. This, and analysis of the structure of the transformations relevant for the Schwinger model will be reported in a forthcoming communication \cite{UsLKFT}.

In ongoing work, we aim to extend these results to QCD and other gauge theories, taking advantage of recent developments in the worldline representation of non-Abelian field theories \cite{Col1, Col2, JO1, JO2, ColTree, CreteProc}. Other future generalisations include analysis of the QED or QCD vertices in covariant gauge and their decomposition into form factors building upon \cite{GluonVertex8,GluonVertex9,GluonVertex12, GluonVertex10} to acquire additional restrictions for analyses of Schwinger-Dyson equations, to the analogous transformations produced by virtual graviton loops along the particle worldlines or to reduced QED as investigated for the propagator in \cite{RQED1}.
\vspace{-1em}
\subsubsection*{Acknowledgments}
\vspace{-1em}
\begin{acknowledgments}
The authors are grateful to Adnan Bashir for a number of useful discussions around the LKFTs in scalar QED and their use in the context of Schwinger-Dyson equations and for recommending various references. They also thank Pietro Dall'Olio for helpful comments. JN and JPE appreciate financial support from CONACyT. JPE also received funding from CIC-UMICH.
\end{acknowledgments}
\clearpage \onecolumngrid
\appendix

\section{Appendix: Photon expectation values}
In the main text we frequently use the properties of the photon propagator, or Green function, defined with respect to a quantum background field, $\bar{A}$, given as $G_{\mu\nu}(x - x^{\prime}; \xi) := \langle \bar{A}_{\mu}(x)\bar{A}_{\nu}(x^{\prime})\rangle_{\xi}$. It is useful to consider the following Euclidean path integral\footnote{We work throughout in $D$-dimensional Euclidean space.} representation of the more general correlation function,
\begin{equation}
	\big\langle \Omega[\bar{A}] \big\rangle_{\bar{A}\, \xi} := \int \mathscr{D}\bar{A}(x) \Omega[\bar{A}] \e^{- \int d^{D}x \big[-\frac{1}{4} \bar{F}_{\mu\nu}\bar{F}^{\mu\nu} - (\partial \cdot \bar{A})^{2}/(2 \xi) \big]}\,,
	\label{eqOmegaPI}
\end{equation}
where $\Omega$ is an arbitrary functional of $\bar{A}$. The functional integral over configurations of $\bar{A}$ is specified with Maxwell action that has been gauge fixed to the linear covariant gauge with parameter $\xi$. In particular, with $\Omega[\bar{A}] = \bar{A}_{\mu}(x)\bar{A}_{\nu}(x')$ we acquire the propagator. It is useful, as we exploit in the main text, to generate insertions of $\bar{A}$ under the path integral by functional differentiation of a source, $J(x)$, so that, for example,
\begin{equation}
	e^{2}G_{\mu\nu}(x - x'; \xi) =\frac{-i\delta}{\delta J^{\mu}(x)}\frac{-i\delta}{\delta J^{\nu}(x')}\mathcal{Z}[J]\Big|_{J = 0}\, ; \qquad \mathcal{Z}[J]:= \int \mathscr{D}\bar{A}(x) \, \e^{- \int d^{D}x \big[-\frac{1}{4} \bar{F}_{\mu\nu}\bar{F}^{\mu\nu} - (\partial \cdot \bar{A})^{2}/(2 \xi)- ie J(x) \cdot \bar{A}(x) \big]}\,.
\end{equation}
Since the path integral in the gauge-fixed partition function, $\mathcal{Z}[J]$, is Gaussian it can be computed exactly in what is an elementary calculation, from which follows the configuration space propagator
\begin{equation}
	G_{\mu\nu}(y; \xi) = \frac{1}{4\pi^{\frac{D}{2}}}\left\{ \frac{1 + \xi}{2}\Gamma\Big[\frac{D}{2} - 1\Big]\frac{\delta_{\mu\nu}}{y^{2}{}^{\frac{D}{2}-1}} + (1 - \xi)\Gamma\Big[\frac{D}{2}\Big] \frac{y_{\mu}y_{\nu}}{y^{2}{}^{\frac{D}{2}}}\right\}.
	\label{eqGConfig}
\end{equation}
This expression follows the general decomposition given in equation (6) of the main text, since the $\xi$ dependent (longitudinal) terms are generated by -- see equation (7) -- differentiation of
\begin{equation}
	\Delta_{D}(y) = -ie^{2}(\mu)\mu^{4-D}\int \frac{d^{D}k}{(2\pi)^{D}} \,  \frac{\e^{-i k \cdot y}}{k^{4}} = -\frac{ie^{2}(\mu)}{16\pi^{\frac{D}{2}}}\Gamma\Big[\frac{D}{2} - 2\Big](\mu y)^{4-D}.
\label{eqDeltaDy}
\end{equation}
\subsection{Fermionic N-point functions}
We write the correlation functions in the first quantised \textit{worldline formalism} of quantum field theory which is equivalent to the standard approach. It can be traced back to early work by Feynman \cite{FeynSc, FeynSp} that was later developed by Strassler \cite{Strassler1, Strassler2} -- for reviews see \cite{ChrisRev, UsRev}. An adequate worldline description of tree level processes has only lately been achieved \cite{fppaper1}, and uses the second order formulation of the Dirac theory \cite{SO1, SO2}. For the propagator (2-point function) this is (c.f. equations (8) and (9) of the main text)
\begin{equation}
	S^{x'x}[A] = 2^{-\frac{D}{2}}\big[m+i\Dsp\big] \symb^{-1} \int_{0}^{\infty} \!dT e^{-m^{2}T}  \!\int_{x(0) = x}^{x(T) = x'}\hspace{-1.5em}\mathscr{D}x\int_{\psi(0) + \psi(T) = 0}\hspace{-3em}\mathscr{D}\psi \, \e^{-\int_{0}^{T}d\tau\, \left[ \frac{\dot{x}^{2}}{4} + \frac{1}{2} \psi \cdot \dot{\psi} +ie \dot{x} \cdot A(x) -ie (\psi + \eta) \cdot F(x) \cdot (\psi + \eta)\right]}\, ,
	\label{eqSWL}
\end{equation}
where the symbol map generates its matrix structure according to equation (9) (we use the convention ${\{\gamma^{\mu}, \gamma^{\nu}\} = -2\eta^{\mu\nu}}$). The key to obtaining the results of the main text is its equation (12), repeated here:
\begin{equation}
	\mathcal{I}[J, M; \xi):= \Big \langle \e^{ie \int d^{D}x J(x)\cdot \bar{A}(x)}\prod_{j=1}^{M}K_{j}^{x'_{\pi(j)}x_{j}}[A^{\gamma}+ \bar{A}]\Big\rangle_{\bar{A}, \,\xi}\,,
\end{equation}
where the expectation value is determined by (\ref{eqOmegaPI}). The worldline action of (\ref{eqSWL}) is, after decomposing ${A = A^{\gamma} + \bar{A}}$ as described in the main text, linear in $\bar{A}$ and the path integral can be determined. Using the path integral representation\footnote{As will be reported in a later communication \cite{UsLKFT}, the calculation is greatly simplified by using a superspace formalism, although we avoid this unnecessary step here at the expense of less compact equations.} of the kernel in equation (8) we find:
\begin{equation}
		\mathcal{I}[J, M; \xi) = \prod_{j=1}^{M}2^{-\frac{D}{2}} \text{symb}^{-1}\int_{0}^{\infty}dT_{j}\, \e^{-m^{2}T_{j}}\int\mathscr{D}x(\tau_{j}) \int \mathscr{D}\psi(\tau_{j})\, \e^{-\sum_{l = 1}^{M}S_{0,\gamma}^{(l)}[x_{l}, \psi_{l}| A^{\gamma}] - \frac{e^{2}}{2} \iint d^{D}y\, d^{D}y' \mathscr{J}(y) \cdot G(y - y'; \xi) \cdot \mathscr{J}(y')}\, ,
		\label{eqI}
\end{equation}
where $S_{0, \gamma}^{(l)}[x_{l}, \psi_{l}| A^{\gamma}]$ comprises the free action for trajectory $l$ and its coupling to the external photons and we have defined a more general current 
\begin{equation}
	\mathscr{J}^{\mu}(y) = J^{\mu}(y) - \sum_{l=1}^{M}\int d\tau_{l}\big[\dot{x}_{l}^{\mu} - 2(\psi^{\mu}_{l}+ \eta^{\mu}_{l})(\psi_{l} + \eta_{l})\cdot \partial \big]\delta^{D}(y - x(\tau_{l}))\,.
	\label{eqJ}
\end{equation}
The crucial observation is that, using (\ref{eqGConfig}), a change in the gauge parameter $\xi \rightarrow \xi + \Delta \xi$ causes a variation in the final term in the exponent of (\ref{eqI}) that can be written in terms of derivatives as
\begin{equation}
	\mathscr{J}(y) \cdot \Delta_{\xi}G \cdot \mathscr{J}(y') = \frac{\Delta \xi}{16 \pi^{\frac{D}{2}}} \Gamma \big[\frac{D}{2} - 2\big] \mathscr{J}(y) \cdot \partial_{y} \mathscr{J}(y') \cdot \partial_{y^{\prime}} \big[(y - y^{\prime})^{2}\big]^{2 - \frac{D}{2}}\, .
	\label{eqDeltaJ}
\end{equation}
Substituting (\ref{eqJ}) and integrating by parts, the Grassmann nature of the variables $\psi_{l}^{\mu}$ and $\eta^{\mu}_{l}$ means that they drop out of (\ref{eqDeltaJ}). Therefore, the terms independent of the source, $J$, give precisely the same scalar factor found in \cite{LKFTWL1, LKFTWL2},
\begin{equation}
	\sum_{k, l = 1}^{M} \frac{\Delta\xi e^{2}}{32 \pi^{\frac{D}{2}}} \Gamma\Big[\frac{D}{2} - 2\Big]\int_{0}^{T_{k}}\hspace{-0.25em}d\tau_{k} \int_{0}^{T_{l}}\hspace{-0.25em}d\tau_{l} \, \partial_{\tau_{k}}\partial_{\tau_{l}}\left[(x_{k} - x_{l})^{2}\right]^{2- \frac{D}{2}} \equiv \sum_{k,l = 1}^{M}\Delta_{\xi}S_{i\pi}^{(k, l)}\,,
\end{equation}
which is the second line of equation (13) in the main text. It corresponds to the transformation of the propagator caused by a change of gauge in the internal photon propagators that couple to the worldline trajectories.  The two terms in (\ref{eqDeltaJ}) involving the source then provide $\Delta_{\xi}I_{M}$ of equation (13), 
\begin{equation}
	\hspace{-1em}\Delta_{\xi}I_{M} = \frac{\Delta \xi e^{2}}{16 \pi^{\frac{D}{2}}}\Gamma\Big[\frac{D}{2} - 2\Big]\hspace{-0.25em} \left(\sum_{i = 1}^{M} \int_{0}^{T_{i}}\hspace{-0.5em}d\tau_{i}\hspace{-0.25em}\int d^{D}x J(x) \cdot \partial_{x} \partial_{\tau_{i}}\big[(x - x_{i})^2\big]^{2-\frac{D}{2}} -\frac{1}{2}\iint\hspace{-0.25em} d^{D}x d^{D}x^{\prime} J(x) \cdot \partial_{x}J(x^{\prime})\cdot \partial_{x^{\prime}}\big[(x - x^{\prime})^{2}\big]^{2 - \frac{D}{2}}\right)\, .
	\label{eqDeltaIm}
\end{equation}
These results imply that $\mathcal{I}$ transforms as (as mentioned in the text, the transformation induced by variation in external photons' gauge is described by the Ward identity)
\begin{equation}
	\hspace{-0.225em} \mathcal{I}[J, M; \xi + \Delta \xi) = \mathcal{I}[J,M; \xi)\e^{-\sum_{k, l=1}^{M}\Delta_{\xi}S_{i\pi}^{(k, l)} + \Delta_{\xi}I_{M}}\,,
	 \label{eqIMsup}
\end{equation}
in agreement with equation (12) above. 

This would be everything were in not for the covariant derivatives in (\ref{eqSWL}), which act through onto the exponent in (\ref{eqIMsup}) -- see equation (14) of the main text. As argued in the text, the derivatives of this exponent are cancelled exactly by additional terms produced by polynomial insertions of $\bar{A}$ from the same covariant derivatives. This cancellation is illustrated for the propagator in equation (15) and here we detail the general case that results in (21). 

First we recall the definition (18) in the main text, 
\begin{equation}
\mathcal{J}[J, K, M; \xi):= \Big\langle \e^{ie \int d^{D}x J(x)\cdot \bar{A}(x)}\prod_{i=1}^{K}\big[m + i \Dsp_{i}\big]\prod_{j=1}^{M}K_{j}^{x'_{\pi(j)}x_{j}}[A^{\gamma}+ \bar{A}]\Big\rangle_{\bar{A}, \,\xi}\,. \label{eqJKsup}
\end{equation}
Defining $\Delta_{\xi}I_{M}^{(2)}$ as the second term of $\Delta_{\xi}I_{M}$ in (\ref{eqDeltaIm}), quadratic in $J$, we prove by induction on $K$ the transformation,	
\begin{equation}
	\mathcal{J}[J, K, M; \xi+ \Delta \xi) = \bigg[\mathcal{J}[J, K, M; \xi)+ \sum_{k = 0}^{K-1} \prod_{l = 1}^{k} \big[m + i {\Dsph}_{l}\big]\big(i \dsp_{k+1}\Delta_{\xi}I_{M}^{(2)} \big) \mathcal{J}^{(k+2)}[J, K,M; \xi) \bigg]	\e^{-\sum_{k, l=1}^{M}\Delta_{\xi}S_{i\pi}^{(k, l)} + \Delta_{\xi}I_{M}}\, ,
	\label{eqDeltaJcal}
\end{equation}
where we defined $\jhat{D}_{\mu} = \partial_{\mu} + \frac{\delta}{\delta J^{\mu}} +ieA^{\gamma}_{\mu}$, whose derivatives \textit{act through} onto everything to their right. The superscript in ${\mathcal{J}^{(k+2)}[J, K, M; \xi)}$ of the second term indicates that the variable $i$ in the definition (\ref{eqJKsup}) runs from $k+2$ to $K$, being ($K-k-1$) values. The $K = 0$  and $K = 1$ cases are already verified by previous calculations. 

As outlined in the main text, for the $K = \kappa$ case we use 
\begin{equation}
\mathcal{J}[J, \kappa,  M; \xi) = \big[m + i\Dsph_{1}\big]\mathcal{J}^{(2)}[J, \kappa, M; \xi) \,,
	\label{eqJkappaSup}
\end{equation}
and the inductive hypothesis, along with 	(compare with equation (16) above)
\begin{equation}
	\frac{\delta}{\delta \Js(x_{i})}\Delta_{\xi}I_{M} - \dsp_{i}\Big[ \sum_{k, l = 1}^{M}\Delta_{\xi}S^{(k, l)}_{i\pi} - \Delta_{\xi}I_{M}\Big] = \dsp_{i}\Delta_{\xi}I_{M}^{(2)}\,,
\end{equation}
provide the gauge variation
\begin{align}
\mathcal{J}[J, \kappa,  M; \xi + \Delta \xi) &=\Big[\mathcal{J}[J, \kappa, M; \xi)+ \big(i\dsp_{1}\Delta_{\xi}I_{M} \big)\mathcal{J}^{(2)}[J, \kappa, M; \xi)\Big]\e^{-\sum_{k, l=1}^{M}\Delta_{\xi}S_{i\pi}^{(k, l)} + \Delta_{\xi}I_{M}}\nonumber \\
&+  \bigg[\sum_{k = 0}^{K-2} \prod_{l = 1}^{k+1} \big[m + i {\Dsph}_{l}\big]\big( i\dsp_{k+2}\Delta_{\xi}I_{M}^{(2)} \big) \mathcal{J}^{(k+3)}[J, \kappa,M; \xi) \bigg]	\e^{-\sum_{k, l=1}^{M}\Delta_{\xi}S_{i\pi}^{(k, l)} + \Delta_{\xi}I_{M}}\, .
\end{align}
Renaming $k \rightarrow k + 1$ in the second line, the second term on the first line supplies the $k = 0$ contribution to the sum on the second and we arrive at the result claimed in (\ref{eqDeltaJcal}) for $K = \kappa$, thus completing the proof.

To see that the additional derivative terms do not contribute when $J = 0$ we note that acting with more than two distinct derivatives (whether they be partial or functional) on $\Delta_{\xi}I_{M}^{(2)}$ annihilates it. Yet we would have to apply two functional derivatives -- on top of the existing partial derivative -- in order for these terms to contribute when $J = 0$. In particular, then, we recover the principal result reported in the main text,
\begin{equation}
	\Big\langle \prod_{i=1}^{K}\big[m + i \Dsp_{i}\big]\prod_{j=1}^{M}K_{j}^{x'_{\pi(j)}x_{j}}[A^{\gamma}+ \bar{A}]\Big\rangle_{\bar{A}, \,\xi + \Delta \xi} =\e^{-\sum_{k, l=1}^{M}\Delta_{\xi}S_{i\pi}^{(k, l)}} \Big\langle \prod_{i=1}^{K}\big[m + i \Dsp_{i}\big]\prod_{j=1}^{M}K_{j}^{x'_{\pi(j)}x_{j}}[A^{\gamma}+ \bar{A}]\Big\rangle_{\bar{A}, \,\xi}\,.
\end{equation}
For $K = M$ we get the configuration space generalised LKF transformation for arbitrary partial amplitudes illustrated in the followng figures. 
\begin{figure}[!htb]\centering
   \begin{minipage}{0.48\textwidth}
     \frame{\includegraphics[width=.7\linewidth]{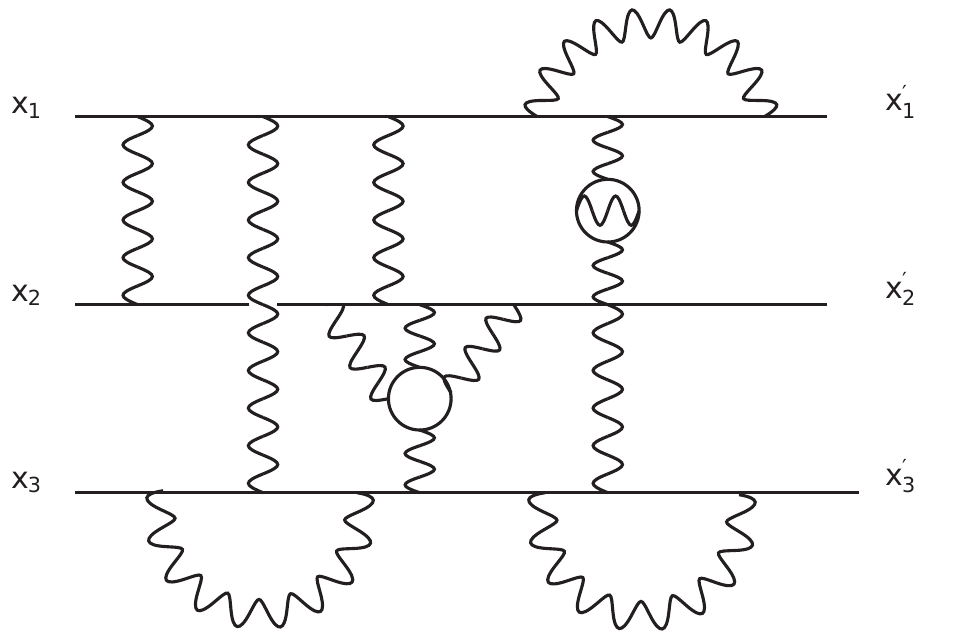}}
     \caption{One contribution to a typical multi-loop correlation function that includes photons attached to internal loops.}\label{Fig:Data1}
   \end{minipage}
   \begin {minipage}{0.48\textwidth}
     \frame{\includegraphics[width=.7\linewidth]{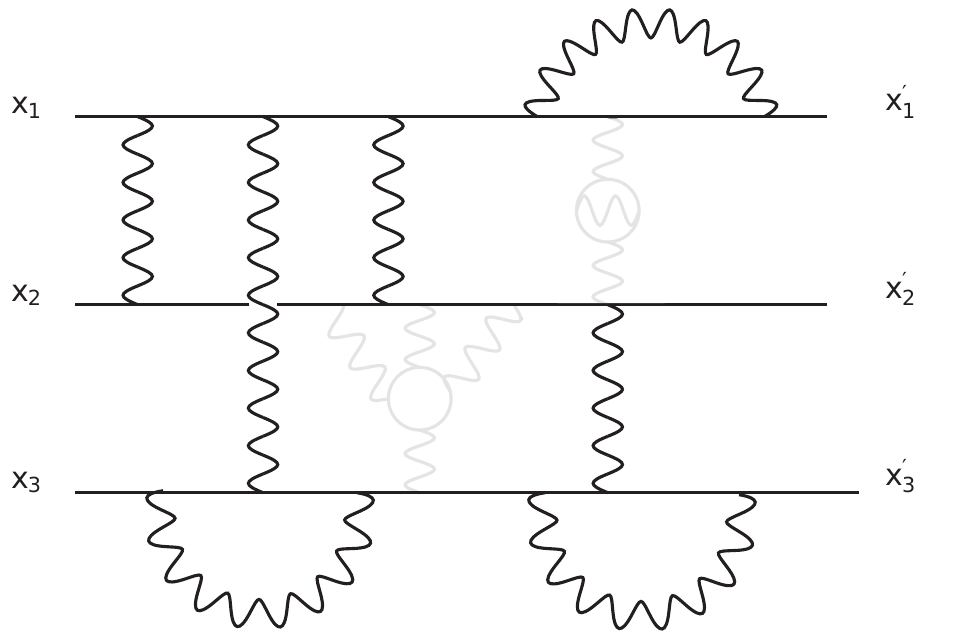}}
     \caption{The LKF transformation is determined by the gauge variation of only the photons attached at both ends to particle lines.}\label{Fig:Data2}
   \end{minipage}
\end{figure}

\bibliography{bibLKFT}
\end{document}